# Statistical Mechanics: A Possible Model for Market-based Electric Power Control

David P. Chassin
*Pacific Northwest National Laboratory*
david.chassin@pnl.gov

**Abstract**

*Statistical mechanics provides a useful analog for understanding the behavior of complex adaptive systems, including electric power markets and the power systems they intend to govern. Market-based control is founded on the conjecture that the regulation of complex systems based on price-mediated strategies (e.g., auctions, markets) results in an optimal allocation of resources and emergent optimal system control. This paper discusses the derivation and some illustrative applications of a first-principles model of market-based system dynamics based on strict analogies to statistical mechanics.*

## 1. Introduction

The idea of controlling a complex engineered system using one or more market-like processes is not new, nor has it always been done with a literal market in the sense that price and quantity are the joint means of determining the allocation of a scarce resource. Indeed, in the early 1980s, so-called contract networks were demonstrated to allocate scarce central processing unit (CPU) time to competing tasks in computers [1]. Nevertheless, the challenges faced by system designers who wish to use market-based control strategies are daunting. Such systems are generally called complex adaptive systems because they exhibit a property that is regarded as a profound obstacle to crafting stable and robust classical control strategies: emergent behavior [2,3,4,5]. This is the property of systems that causes them to exhibit global behaviors not anticipated by the control strategy. For example, building heating, ventilation, and air-conditioning (HVAC) systems exhibit a phenomenon called global hunting, an accidental artifact of the control strategies employed to govern the thermodynamic process [6]. Global hunting exhibits itself as a unexpectedly prolonged cycling between two or more local minima.

Many systems, in molecular biology [7], power engineering [8], sociology [9], military command / control / communications / intelligence ($C^3I$), and air traffic control, exhibit such unanticipated emergent behavior and are thus candidates for membership in the class of systems characterized by complex adaptive behavior with the potential for unanticipated robustness degradation [10,11]. However, exploiting the emergent behavior of certain systems [12] by design has been the subject of discussions with respect to building controls [13] and using price as the only signal for electric power systems control [14]. The challenge to exploiting such phenomena is predicting the emergent behavior of complex systems. This is fraught with analytic difficulties, many of which are discussed by Atmanspacher and Wiedenmann [15].

It is with the object of understanding the relationship between the rules governing the behavior of machines and the emergent econophysical behaviors of the systems that we seek a theory of control based on statistical mechanics. Tsallis et al. [16] outlined how one might apply non-extensive statistical mechanics to the question of stability and robustness. Donangelo and Sneppen [17] went further to discuss the dynamics of value exchange in complex systems. While the full development of a theory of complex system behavior is obviously beyond the scope of a single paper, the first step in this endeavor for energy system control is to devise a rigorous model of the aggregate properties of a system based on those of individual agent-machines. *A priori*, it is by no means certain that the analogy to statistical mechanics should hold strictly. But insofar as it does, we can proceed with the derivation of such properties as are consistent with it. Ultimately we can conclude that this analogy is true only if we find no inconsistencies and the aggregate properties we develop usefully elucidate the behavior of systems.

A machine such as that shown in Figure 1 is conserving [18] so that over time the total value of products $Q$ acquired, delivered and lost (e.g., in kWh) by the thermal process is equivalent to the costs $C$ received, paid, invested and withdrawn by the control process (e.g., in \$/h) at the prevailing prices. While the state $x$ of the thermodynamic process is theoretically continuous (e.g., kW), the state $n$ of the control process for loads is often discrete (e.g., *on*, *off*) particularly for smaller, more numerous and indistinguishable machines. We expect that having a formal model of conserving components makes it possible to postulate laws for systems comprised of many such interacting components in strict analogy to statistical mechanics. Ultimately we hope to develop methods of control for complex adaptive systems designed so that these aggregate system properties will be experimentally verifiable and have probabilities that may be approximated in closed analytic form.

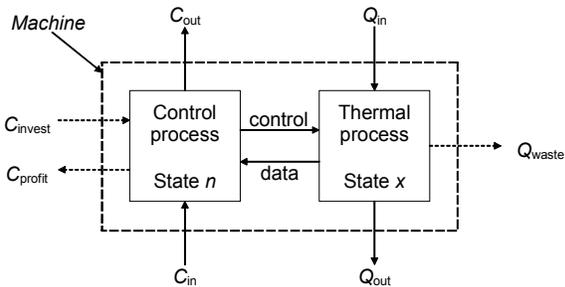

Figure 1: The abstract transactive machine.

We expect the price of products exchanged in such systems to be an important factor in the system's dynamics. However, we do not expect the particular features or rules of market clearing to be relevant, insofar as most price determination methods (e.g., auctions, bilateral negotiations, or price controls) are ubiquitous within the system. Indeed, for systems where several price determination methods are extant, it should be the subject of future discussion to explore how the dominance of the most globally efficient method can be presumed. However, we need not demonstrate this to pursue the present discussion.

Finally, we do not suggest that this approach is by any means exclusive of previous efforts to gain insights into the interactions between power systems and markets. Indeed, Stoft's comprehensive work on the design of markets for electricity is an excellent summary of those [19]. However, that work illustrates the complexity of the existing discussion. For those who seek a more coherent and elegant description of complex systems, these discussions are discomforting and ultimately do not shed enough light on the underlying principles extant.

## 2. Statistical mechanics of econophysics

Today's electric power systems are generally composed of only a few thousand entities that are economically interacting (e.g., generators, transmission operators, distribution companies, major customers), relegating the vast majority of customer loads to a relatively marginal role, having only electromechanical significance with respect to shorter-term phenomena. However, future systems are envisioned in which demand response plays a much more significant role, and in which potentially many millions of loads are economically involved as well. For example, it is expected that price-responsive control strategies in appliances such as air-conditioners and refrigerators will become more commonplace. This evolution increases the complexity of the modeling problem by many orders of magnitude.

Efforts are already under way to develop simulation systems that will enable the study of this much more complex type of system [20], however many interesting phenomena remain out of reach of these models [21].

We propose to build a robust model for the theoretical treatment of the joint economic, logical and physical behavior of an ideal power system composed of many millions of potentially interacting machines. First we need to adopt uniform terminology to deal with the incompatible definitions of terms such as *rate*. Rate denotes two very distinct concepts in economics and in physics. Our purpose is not to supplant the well-established definitions in each domain, but to avoid the confusion caused by the semantic differences conferred by them to identical terms. We will occasionally return to the subject of terminology as the situation demands.

We begin with the notion of system *state*. State simply refers to a configuration of a system or machine. By configuration we mean a combination of the states of the components of a system or machine. A *stationary state* refers to the property of a system wherein the probabilities of certain states occurring remain constant over time. When we speak of *discrete states*, we refer to the condition of a system wherein all the physical properties of the system we consider are denumerable (enumerated but possibly infinite). We should also keep in mind that the system being considered might be composed of a single machine or many machines. Usually we are concerned with systems comprised of many machines, but sometimes we will consider a system as a single machine. Such a reduction will serve our purposes without detracting from the result. Continuous states, however, may not sustain an actual physical solution; real systems are often described only as linear combinations of these states (integrals over some parameter). We assume that each discrete system state has a definite value in an economic sense, but many discrete states may share the same value.

*Value* and *energy* are considered interchangeable abstractions when assessing the equilibrium dynamics of systems. Where we speak of systems of *constant energy* in physics, we will speak of systems of *constant value* in econophysics. Indeed, a crucial assumption is that there is equivalence between energy and value, which is established according to the current market price. Therefore, we may speak of the energy or value of states interchangeably, but we will give strong preference to the term *value* in recognition of the expanded semantic required by an econophysical treatment of system dynamics.

In addition we recognize that systems have various values with varying likelihood, depending on conditions. The *degeneracy* of a system value level refers to the number of discrete states having that value or in a narrow range of values near it. Thus, it is the value and not the discrete state that has a degeneracy. The definition of the degeneracy of a value level is quite distinct from the physical degeneracy of states and is principally understood with respect to a method of measurement of value. What might look like a single degeneracy according to one method, can be split into multiple levels using another method. This does not detract from the validity of either method. Indeed it is a useful peculiarity of this treatment of econophysics that we may choose one method or another as appropriate to the problem being considered.

The *total value* of a system (denoted $U$) includes both the *potential value* of each machine and the *value of activity*, which takes into account all mutual interactions. Thus, the value of a system of two or more machines cannot be described exactly as the excitation value (i.e., the non-quiescent value) of each machine under the influence of another, in spite of the fact that this is often a very good approximation to describe the low-activity levels of systems. We must also consider the potential value inherent in the configuration of the machines' states. However, it remains necessary to distinguish between the discrete states of the system of $N$ machines and those of each machine, even when referring to a system of only one machine. Therefore, when referring to the state of a single machine, we will speak of its *mode*, the value of which we denote $\varepsilon$.

Because we are ultimately concerned with properties of econophysical systems of many different types, we will describe the statistical properties of systems of $N$ machines for which we know the set of values $\varepsilon_l(N)$, which denotes the value of the discrete state $l$ of a $N$–machine system. The Greek letter *epsilon* is used interchangeably with energy only conceptually in the context of a statistical interpretation of conserving systems. The index $l$ denotes the discrete state number. It can be assigned in any way, with the only caveat that no two discrete states with differing values may have the same index.

We will begin by studying the properties of a simple system for which the values $\varepsilon_l(N)$ may be calculated exactly. The conclusions obtained from this model system will be assumed provisionally to apply to all econophysical systems. At this point it is a significant conceptual leap. However, the strict adherence to the analogy with thermal physics allows us to reasonably expect results consistent with experiments properly put forth. We should expect this discussion to provide insight into the nature and scope of such experiments.

## 2.1. States of a system

We gain confidence in this method of considering a many-machine econophysical control problem by solving a simple system exactly. The proposed system is abstract, however, there are examples of machines that exist as special cases of this abstract machine. For instance, a building equipped with sufficient on-site generation to fully satisfy its own load can be simplified according to this model. Such a building may either produce all its power from the on-site source (in which case it is not in the system), produce more power that it needs (in which case it is a seller), or consume more power than it produces (in which case it is a buyer).

Given a system of $N$ identical and independent machines, each producing (denoted by $S$) or consuming (denoted by $B$) a fixed quantity of power, $Q$ on a line, we specify the states of the system as the possible arrangements of the machines, of which there are $2^N$. We exclude machines that are either completely off or self-supplying because they do not exist for this simple model of production/consumption. For example, the arrangement *BSSBSBSB* shown in Figure 2 is one possible state of a system of 8 machines having 256 ($2^8$) possible states.

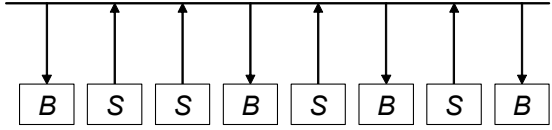

Figure 2: One possible equi-partitioned state of a system of 8 abstract machines on a line.

Of the 256 distinct states of the system, we count only the 9 distinguishable states because we are ultimately concerned with knowing only the number of machines having state *B* or *S*, and not the specific state of each machine. Ultimately we seek an expression for the number of distinguishable states of a system of *N* machines in which $\frac{1}{2}N \pm m$ are consumers and producers, respectively and the difference $2m =$ (# of *producers*) – (# *of consumers*) is called the *excess product*.

To illustrate the significance of this, consider that in an *N*-machine system there is only one state in which all the machines are producers and none are consumers. But there are *N* states for which only one is a consumer and (*N*–1) are producers, and there are $\frac{1}{2}N(N–1)$ for which two are consumers and (*N*–2) are producers, etc. The most likely state of a system of periodic loads having equal probability of being producers and consumers is $\frac{1}{2}N$ of each. We seek an analytic expression for the number of states with $\frac{1}{2}N + m$ producers and $\frac{1}{2}N - m$ consumers. (Note that it is convenient to assume that *N* is even in spite of the fact that it may not be. We are generally concerned with values of *N* that are very large, so it does make a significant difference if *N* is odd.) We may symbolically write the *N*-factor product (the italicized digits are simply symbols for the states of individual machines) as $(B + S)^N$, which we may expand using the binomial theorem so that it is given by

$$\sum_{m=-\frac{1}{2}N}^{\frac{1}{2}N} \frac{N!}{\left(\frac{1}{2}N - m\right)!\left(\frac{1}{2}N + m\right)!} B^{\left(\frac{1}{2}N-m\right)} S^{\left(\frac{1}{2}N+m\right)}.$$

The coefficient of $B^{(\frac{1}{2}N-m)}S^{(\frac{1}{2}N+m)}$ is denoted by $g(N,m)$ and is the state degeneracy function (Kittel and Kroemer refer to this function as the *multiplicity function* in the 1980 edition [22] but Kittel uses the term degeneracy function in his 1969 edition.) It is a count of the number of system configurations for which 2*m* more of *N* loads are selling than buying. We can show that the degeneracy function is approximated as

$$g(N,m) \cong g(N,0)e^{-2m^2/N}$$

where, using Stirling's approximation we find for that for large *N*

$$g(N,0) = \frac{N!}{\left(\frac{N}{2}!\right)^2} \approx 2^N \sqrt{\frac{2}{\pi N}}.$$

Chapter 1 of Kittel contains a complete discussion of the thermodynamic derivation of the degeneracy function for binary model systems such as this. Indeed the mathematical treatment of the subject in this paper is largely based on Kittel's examination of various thermodynamic models and in many cases follows precisely the same reasoning and notation.

## 2.2. Entropy and the value activity

Now we may develop a basis for a definition of *entropy* and measure for the value of *activity* in a system, stemming from the interaction of machines. In the process we will rigorously define both entropy and activity. But before proceeding we will establish some assumptions that we must make. The computation of the entropy and activity of a system is based on the assumption that we are considering a closed system in which any stationary discrete state is equally likely if it is an *accessible state*. By closed system we mean a system in which the total value, the total number of machines, and the total number of states all remain constant. A state is accessible if its properties are compatible with the specifications of the system. This will usually mean that its value must be within a given range of the values specified for the system, and the number of machines represented by the state must be equal to the number of machines in the specification of the system. Basically, we will treat all discrete states that satisfy the value constraints as accessible unless the specification of the system and the time scale of the measurement somehow exclude them.

## 2.3. Two systems in contact

Our purpose is to quantify entropy and the value of activity in systems. To accomplish this we must determine the number of *accessible states* of a system, the logarithm of which is the entropy. Entropy is the key to understanding the dynamic properties of econophysical systems. Using this, we may consider what we learn from analyzing the entropy of two systems that come into contact with each other.

Let us therefore consider the condition when two systems, 1 and 2, of independent machines come into contact with each other, forming a combined complex within which value may flow. We permit value to flow, but we will not permit the machines themselves to move from one system to the other, so we are considering a condition wherein only *transactive contact* prevails. The excess products, $2m_1$ and $2m_2$ of each system may be different. The actual exchange of value may take place by any means across the interface of the two systems. We will keep the number of respective machines $N_1$ and $N_2$ constant, but we should expect the excess product to change. We express the sum of the excess product for the complex as $2m$ and we have

$$m = m_1 + m_2$$

The total value $U(m)$ of the complex with excess product $2m$ is

$$U(m) = U_1(m_1) + U_2(m_2) \quad (1)$$

and the number of machines is

$$N = N_1 + N_2.$$

We will assume the value levels are the same in both systems, so that the value obtained by system 1 when a machine changes from buyer to seller can be taken up by a complementary reversal of the mode of a machine in system 2.

To determine the degeneracy function $g(N,m)$ of the complex, we must reason that the configuration of the combined system is based on the enumeration of states where each state of system 1 can occur in combination with every accessible state of system 2. Thus, the degeneracy function of the complex is based on the product of the degeneracy functions of the contributing systems, $g_1(N_1,m_1)g_2(N_2,m_2)$, and because $m = m_1 + m_2$ we find that

$$g(N,m) = \sum_{m_1} g_1(N_1, m_1) g_1(N_2, m - m_1) \quad (2)$$

is the number of accessible states of the combined system. At this point we reason that a product of the form $g_1(N_1,m_1)g_2(N_2,m-m_1)$ will have a maximum for some value of $m_1$, denoted $\hat{m}_1$ for which the number of states in the *most probable configuration* is

$$g_1(N_1, \hat{m}_1) g_2(N_2, m - \hat{m}_1).$$

When the number of machines of the two systems is very large, we can show that the maximum will have a very small variance with respect to changes in $m_1$. This means that a relatively small range of $m$'s near $\hat{m}$ (hence a relatively small number of configurations) will dominate the statistical properties of the complex. We suppose that this is also a property of small systems with precisely computable solutions and is a general property of systems of any size. However, we will assume that at least one of the systems considered in any given problem is sufficiently large that this property holds, and we will call such systems *reservoirs*.

Therefore, let us consider a system of machines as previously discussed. We use the distribution functions for $g_1(N_1,m_1)$ and $g_2(N_2,m_2)$, as shown, above to form the product

$$g_1(N_1, m_1) g_2(N_2, m_2)$$

which is equivalent to

$$g_1(0) g_2(0) e^{\left(-\frac{2m_1^2}{N_1} - \frac{2m_2^2}{N_2}\right)} \quad (3)$$

where $g_1(0)$ and $g_2(0)$ denote $g_1(N_1,0)$ and $g_2(N_2,0)$, respectively. We use these degeneracies of the systems as a reference because the degeneracy for $m = 0$ has a special significance. We may replace $m_2$ with $m - m_1$, thus giving us

$$g_1(0) g_2(0) e^{\left(-\frac{2m_1^2}{N_1} - \frac{2(m-m_1)^2}{N_2}\right)}$$

for which we seek the maximum value as a function of $m_1$ which is the same as the maximum of

$$\log[g_1(0) g_2(0)] - \frac{2m_1^2}{N_1} - \frac{2(m-m_1)^2}{N_1} \quad (4)$$

The maximum is found when the first derivative with respect to $m_1$ is zero, as given by

$$-\frac{4m_1}{N_1} + \frac{4(m-m_1)}{N_2} = 0$$

and when the second derivative with respect to $m_1$ is negative, as given by

$$-4\left(\frac{1}{N_1} + \frac{1}{N_2}\right) < 0$$

which is true for all $N$.

So the condition that satisfies the most probable configuration of the complex is given by

$$\frac{m_1}{N_1} = \frac{m - m_1}{N_2} = \frac{m_2}{N_2},$$

which is to say that the complex is at equilibrium when the fractional excess products of the two systems are equal. So if $\hat{m}_1$ and $\hat{m}_2$ denote the values of $m_1$ and $m_2$ at the maximum, then the equilibrium condition is met when

$$\frac{\hat{m}_1}{N_1} = \frac{\hat{m}_2}{N_2}. \tag{5}$$

Therefore we can evaluate the product $g_1g_2$ at the maximum defined as

$$(g_1g_2)_{max} \equiv g_1(\hat{m}_1)g_2(m - \hat{m}_1)$$

by substitution of (5) into (3) such that

$$(g_1g_2)_{max} = g_1(0)g_2(0)e^{-2m^2/N}. \tag{6}$$

As to the question of how sharp the maximum of $g_1g_2$ is, consider a deviation $\delta$ of $m_1$ and $m_2$ from their respective values at the maximum, such that

$$m_1 = \hat{m}_1 + \delta \quad ; \quad m_2 = \hat{m}_2 - \delta$$

which we square and substitute in (3) and use (6) to obtain the number of states

$$g_1g_2 = (g_1g_2)_{max} e^{\left(-\frac{4\hat{m}_1\delta}{N_1} - \frac{2\delta^2}{N_1} + \frac{4\hat{m}_2\delta}{N_2} - \frac{2\delta^2}{N_2}\right)}.$$

From (5) we know that $\hat{m}_1/N_1 = \hat{m}_2/N_2$, and therefore the number of states in configuration of deviation $\delta$ is given by

$$g_1(N_1, \hat{m}_1 + \delta)g_2(N_2, \hat{m}_2 - \delta)$$
$$= (g_1g_2)_{max} e^{-2\delta^2\left(\frac{1}{N_1} + \frac{1}{N_2}\right)} \tag{7}$$

and for $N_2 \gg N_1$

$$(g_1g_2)_{max} e^{-2\delta^2\left(\frac{1}{N_1} + \frac{1}{N_2}\right)} \approx (g_1g_2)_{max} e^{-\frac{2\delta^2}{N_1}}. \tag{8}$$

The standard deviation is approximately $\sqrt{N_1}$ and the relative width of the peak is roughly $1/\sqrt{N_1}$. A system with 10,000 units would have a relative width of the peak of 1/100, which is a very sharp peak. So while the fractional deviations of small systems can be substantial, those of large systems are not. Therefore, the properties of small systems in contact with large ones are defined as those of the large system.

In reality, power systems will deviate from equilibrium for many reasons. However, (7) quantifies the minimum deviation stemming solely from the randomness of the machines.

We can generalize the result (2) for any two arbitrary systems by incorporating an extension drawn from (1), according to which we obtain a generalized result for the degeneracy $g(N,U)$ of a complex:

$$g(N,U) = \sum_{U_1} g_1(N_1,U_1) g_2(N_2,U-U_1) \tag{9}$$

where the sum is taken over all the values of $U_1 \leq U$. According to this generalized definition, $g_1(N_1,U_1)$ is the number of accessible states of system 1 at the value $U_1$; a configuration of the complex is specified by the values $U_1$ and $U_2$; the number of accessible states of the complex is given by $g_1(N_1,U_1)g_2(N_2,U-U_1)$; and the sum over all configurations gives $g(N,U)$. Because we seek the largest term of the sum in (9), it is necessary that the differential be zero for an infinitesimal exchange of value, because the most probable configuration is at an extremum of the degeneracy function $g$. Thus,

$$\begin{cases} dg = \left(\frac{\partial g_1}{\partial U_1}\right) g_2 dU_1 + g_1 \left(\frac{\partial g_2}{\partial U_2}\right) dU_2 \\ dU_1 + dU_2 = 0 \end{cases}$$

is satisfied for the most probable configuration of the complex. If we divide this result by $g_1g_2$ and substitute $-dU_2$ for $dU_1$, we find that

$$\frac{1}{g_1}\left(\frac{\partial g_1}{\partial U_1}\right)_{N_1} = \frac{1}{g_2}\left(\frac{\partial g_2}{\partial U_2}\right)_{N_2}.$$

We use the fact that $d/dx \log y = 1/y \, dy/dx$ to rewrite this expression as

$$\left(\frac{\partial \log g_1}{\partial U_1}\right)_{N_1} = \left(\frac{\partial \log g_2}{\partial U_2}\right)_{N_2}. \tag{10}$$

The dependence of the systems' value $U_i$ on the number of accessible states $g_i$ determines the most probable configuration of the complex.

## 3. Applications of entropy and activity

From now on we will say that two systems are at equilibrium with each other when the combined complex is in the most probable configuration, i.e., that configuration at which the number of accessible states is a maximum. We have seen that (10) describes the equilibrium condition for the complex, and because $g$ can be very large, we define the entropy as

$$\sigma(N,U) = \log g(N,U) \tag{11}$$

i.e., the logarithm of the number of states accessible to a system of $N$ machines at the value $U$. Thus we may rewrite (10) as

$$\left(\frac{\partial \sigma_1}{\partial U_1}\right)_{N_1} = \left(\frac{\partial \sigma_2}{\partial U_2}\right)_{N_2}.$$

It is from this expression that we come to understand the *value of activity* such that two systems in equilibrium have the same value for this quantity. Thus we are led to define the activity $\tau$ by the relation

$$\frac{1}{\tau} = \left(\frac{\partial \sigma}{\partial U}\right)_N. \quad (12)$$

The value of activity of a system is the change in value with respect to the change in entropy of a system with a constant number of machines. Formally we should refer to $\tau$ as the *activity*. Because the entropy $\sigma$ is dimensionless, $\tau$ has the same unit as the value $U$, which is typically that of currency.

So what determines the flow of value from one system to another? It turns out that the concept of the value of activity plays a crucial role. The direction of the flow is not simply a matter of which system has more total value because the two systems can have quite different size and configurations. Rather it is determined by which system has greater activity, and value will flow from the system with greater activity to the one with less. It is convenient to define $1/\tau$ rather than just $\tau$, as equal to $(\partial \sigma/\partial U)_N$ because it allows our definition to be consistent with the notion that value flows from systems of higher activity to systems of lower activity.

### 3.1. Deviations of systems from equilibrium

To get a sense of the impact of Equation (7), we consider two simple equal-sized systems comprising altogether $N$ residences, each of which is equipped with a single 1-kW on-site power source and a single 1-kW load. The two systems are connected together forming a complex the scale of a national power grid, so that $N_1 = N_2 \approx 2 \times 10^8$. A $\delta = 10{,}000$ (a fractional deviation $\delta/N_1$ of $1.0 \times 10^{-4}$) corresponds to a reduction of $g_{max}$ of approximately $1.8 \times 10^{-2}$. If the cycling time of the machines is about once per hour, such a deviation can be expected to occur about every two days. If each system has a total capacity of about 100 GW, this deviation represents roughly 10 MW. However, a deviation only 150% greater and equivalent to 25 MW, reduces $g_{max}$ by $1.4 \times 10^{-11}$ and is estimated to occur about once every 8 million years, i.e., practically never (see Table 1).

Table 1. Occurrence rate of equilibrium deviations

| MW Deviation | 10 | 15 | 20 | 25 |
|---|---|---|---|---|
| % Deviation | 0.01% | 0.015% | 0.020% | 0.025% |
| $g_{max}$ Reduction | $2 \times 10^{-2}$ | $1 \times 10^{-4}$ | $1 \times 10^{-7}$ | $1 \times 10^{-11}$ |
| Time to event | | | | |
| Days | 2.3 | 340 | | |
| Years | | 0.92 | 1000 | $8.2 \times 10^6$ |

### 3.2. Additivity of entropy

Two closed systems when not in contact have $g_1$ and $g_2$ accessible states with entropy $\sigma_1 = \log g_1$ and $\sigma_2 = \log g_2$, respectively. The combined system has $g_1 g_2$ accessible states so the combined system has $\sigma = \log(g_1 g_2) = \log g_1 + \log g_2 = \sigma_1 + \sigma_2$. In general we should expect to find the entropy of a combined system is the sum of the entropies of the separate systems for $g_1$ and $g_2$ constant. However, it is reasonable to expect that if each system is at its lowest value state, then the number of accessible states might change when they come into contact, so that the entropy is not always additive. Investigation of these departures from non-additivity of entropy will be deferred to future discussions of perturbations and transients.

### 3.3. Increase of value with activity

Consider the spontaneous exchange of value $\delta U_1 = -\delta U_2$ between two systems in transactive contact. To second order we have

$$\begin{cases} \delta\sigma_1 = \dfrac{\partial \sigma_1}{\partial U_1}\delta U_1 + \dfrac{1}{2}\dfrac{\partial^2 \sigma_1}{\partial U_1^2}(\delta U_1)^2 \\ \delta\sigma_2 = \dfrac{\partial \sigma_2}{\partial U_2}\delta U_2 + \dfrac{1}{2}\dfrac{\partial^2 \sigma_2}{\partial U_2^2}(\delta U_2)^2 \end{cases}$$

and because $\partial\sigma_1/\partial U_1 = \partial\sigma_2/\partial U_2$ when two systems have the same activity, the total entropy change to second order in $\delta U$ is

$$\delta\sigma = \delta\sigma_1 + \delta\sigma_2 = \frac{1}{2}(\delta U)^2 \left(\frac{\partial^2 \sigma_1}{\partial U_1^2} + \frac{\partial^2 \sigma_2}{\partial U_2^2}\right).$$

The definition of activity allows us to express the second derivative as

$$\frac{\partial^2 \sigma_1}{\partial U_1^2} = \frac{\partial}{\partial U_1}\left(\frac{\partial \sigma_1}{\partial U_1}\right) = \frac{\partial}{\partial U_1}\left(\frac{1}{\tau_1}\right) = -\frac{1}{\tau_1^2}\frac{\partial \tau_1}{\partial U_1}$$

and because $\tau_1 = \tau_2$ we find

$$\delta\sigma = -\frac{(\delta U)^2}{2\tau^2}\left(\frac{\partial\tau_1}{\partial U_1} + \frac{\partial\tau_2}{\partial U_2}\right).$$

The entropy $\sigma$ is maximal at equilibrium, thus the entropy change $\delta\sigma$ must be negative for any finite exchange of value as a result of a fluctuation from equilibrium. This can only be satisfied for each system with respect to a constant number $N$ of machines when the condition

$$\left(\frac{\partial U}{\partial \tau}\right)_N > 0$$

is true for each system. This tells us that the value of a system increases when the activity increases, an intuitive conclusion that conforms to our expectations for such a property of systems.

## 4. Potential value

The consideration of what occurs when two systems are allowed to exchange value led us to natural definitions of entropy and activity. However, we have yet to consider the consequence of allowing machines to also move from one system to another when they are in contact.

Given the constraints

$$U = U_1 + U_2 \quad ; \quad N = N_1 + N_2, \tag{13}$$

we can express the most probable configuration of the complex as that for which the number of accessible states is a maximum. This occurs when the product of the number of accessible states of the separate systems, expressed by $g_1(N_1,U_1)g_2(N-N_1,U-U_1)$, is at its maximum with respect to $U_1$ and $N_1$. For this condition to be met we must have

$$\begin{aligned}&\left[\left(\frac{\partial g_1}{\partial N_1}\right)_{U_1} dN_1 + \left(\frac{\partial g_1}{\partial U_1}\right)_{N_1} dU_1\right]\\&+ \left[\left(\frac{\partial g_2}{\partial N_2}\right)_{U_2} dN_2 + \left(\frac{\partial g_2}{\partial U_2}\right)_{N_2} dU_2\right] = 0\end{aligned} \tag{14}$$

We know from (13) that

$$\begin{cases} dN_2 = d(N - N_1) = -dN_1 \\ dU_2 = d(U - U_1) = -dU_1 \end{cases}. \tag{15}$$

As before, we divide (14) by $g_1g_2$ and use (15) to obtain

$$d(g_1 g_2) = \left[\frac{1}{g_1}\left(\frac{\partial g_1}{\partial N_1}\right)_{U_1} - \frac{1}{g_2}\left(\frac{\partial g_2}{\partial N_2}\right)_{U_1}\right] dN_1$$
$$+ \left[\frac{1}{g_1}\left(\frac{\partial g_1}{\partial U_1}\right)_{N_1} - \frac{1}{g_2}\left(\frac{\partial g_2}{\partial U_2}\right)_{N_1}\right] dU_1 = 0$$

as the condition for the complex to be at equilibrium. If we use the previous definition of entropy from (11), we know $\sigma = \sigma_1 + \sigma_2$ and we find

$$d\sigma = \left[\left(\frac{\partial \sigma_1}{\partial N_1}\right)_{U_1} - \left(\frac{\partial \sigma_2}{\partial N_2}\right)_{U_1}\right] dN_1$$
$$+ \left[\left(\frac{\partial \sigma_1}{\partial U_1}\right)_{N_1} - \left(\frac{\partial \sigma_2}{\partial U_2}\right)_{N_1}\right] dU_1 = 0.$$

The two terms in the square brackets must be zero for the equilibrium condition to be met, therefore we must have

$$\left(\frac{\partial \sigma_1}{\partial U_1}\right)_{N_1} = \left(\frac{\partial \sigma_2}{\partial U_2}\right)_{N_2} ; \left(\frac{\partial \sigma_1}{\partial N_1}\right)_{U_1} = \left(\frac{\partial \sigma_2}{\partial N_2}\right)_{U_2}.$$

The first condition is a familiar one; it is the one that led to the definition of equilibrium of activity only, or $\tau_1 = \tau_2$. However, the second condition is new and to discuss it further, we must introduce a new concept called *potential value*, denoted $\mu$. Thus we will define potential value by

$$-\frac{\mu}{\tau} \equiv \left(\frac{\partial \sigma}{\partial N}\right)_U. \tag{16}$$

The analogy to electrochemical potential in thermal physics is clear. The potential value is related to the fractional change entropy with respect to a change in the number of machines. For two systems having the same activity $\tau$ we have a more general equilibrium condition that is now

$$-\frac{\mu_1}{\tau} = -\frac{\mu_2}{\tau}$$

or more simply

$$\mu_1 = \mu_2.$$

Thus two systems that can exchange both product and machines are said to be in equilibrium when the value of activity and the potential value are the same for each system, respectively. As before, we have chosen the sign of $\mu$ such that the flow of value goes from the system with greater potential value to the one with lesser potential value.

## 4.1. Potential value of a system in a market

From (4) we know the entropy of a system can be given with respect to the degeneracy $g(N,0)$ of the system at equilibrium. Thus

$$\sigma(N,m) = \log g(N,0) - \frac{2m^2}{N}$$

for $|m|/N \ll 1$. If $Q$ is the amount of power (in kW) used for the contract period $T$ (in hours) at the price $P$ (in \$/kWh), the value $U$ obtained from the system when in contact with the market is given by

$$U = -QTP.$$

We know the quantity $QT/2m$ is the fundamental quantity of interest; it is the unit value $\varepsilon$ from each machine per transaction in a system with excess product $2m$. So we find that

$$U = -2m\varepsilon P$$

Squaring this expression and rearranging we find

$$m^2 = \frac{U^2}{4\varepsilon^2 P^2}$$

so that the entropy can now be stated in terms of $U$:

$$\sigma(N,U) = \sigma(N,0) - \frac{U^2}{2\varepsilon^2 P^2 N}.$$

From the definition of potential value we have

$$-\frac{\mu}{\tau} = \left(\frac{\partial \sigma(N,U)}{\partial N}\right)$$

from which we find that

$$-\frac{\mu}{\tau} = -\frac{\mu(0)}{\tau} + \frac{U^2}{2\varepsilon^2 P^2 N}$$

where $\mu(0)$ is the potential value evaluated at zero value of activity.

According to the definition of activity we find that

$$\frac{1}{\tau} = \left(\frac{\partial \sigma}{\partial U}\right)_N = -\frac{U}{\varepsilon^2 P^2 N} \qquad (17)$$

so that the quantity

$$-\frac{\mu}{\tau} = -\frac{\mu(0)}{\tau} + \frac{\varepsilon^2 P^2}{2\tau^2}$$

or

$$\mu(\tau, P) = \mu(0) - \frac{\varepsilon^2 P^2}{2\tau}. \qquad (18)$$

The potential value of a system decreases as the price of power increases. This may seem counterintuitive at first, but it is much easier to recognize the origin and significance of this statement if we substitute $\tau$ from (17) into (18) and rearrange. We then find that

$$-\mu(0) = -\mu(\tau, P) + \frac{U}{2N}.$$

This result confirms the existence of a quantitative relationship between the value of activity in a system and the potential value of individual machines given a certain market price. We recognize that the price $P$ is increasing the order of the system from that at $-\mu(0)$ to that at $-\mu(\tau,P)$. (This is precisely analogous to a magnetic field that changes the chemical potential of a system of atoms, ordering the spins of the atoms, thereby decreasing the entropy of the system, as described in Kittel and Kroemer [22], page 52.) As expected, this change is equivalent to the exchange of product at the per-unit value $U/2N$. Note that it is maximal only with the full participation of all $N$ machines in the system and limiting the action of one or more machines will reduce the aggregate system value, even though it may benefit a few machines by increasing their trading opportunities.

## 5. Conclusion

In this paper we have elucidated the fundamental elements of a statistical mechanical interpretation of market-based power system econophysics. We have described the basic framework for a treatment of the equilibrium dynamics for systems controlled principally by price-based machine-machine interactions, the so-called transactive control paradigm of electric power systems. We have seen that this treatment builds on a conception of machine state as a denumerated quantity, and that notions of entropy, potential value, and value of activity may be derived formally for systems of these machines. We have found no inconsistencies in this derivation and the results are useful in the sense that they permit us to better understand certain aggregate properties of systems of simple agent-machines. We have demonstrated simple applications of these concepts that illustrate the clarity and rigor of this formalism as applied to power markets and power systems in general.

## 6. Acknowledgments

The author wishes to thank Dr. Jeffry V. Mallow and Dr. Asim Gangopadhyaya of Loyola University Chicago Department of Physics for their insight and advice. This work was supported by Pacific Northwest National Laboratory and the U.S. Department of Energy Office of Electricity Transmission and Distribution. Pacific Northwest National Laboratory is operated for the U.S. Department of Energy by Battelle Memorial Institute under contract DE-AC06-76RL01830.